\documentclass{article}
\usepackage{graphicx}
\usepackage{amssymb}
\usepackage{epstopdf}
\usepackage{amsmath}
\usepackage{natbib}
\usepackage{subfigure}

\begin{document}

{\bf Deep-water gravity waves: theoretical estimating of wave parameters} 
\bigskip

I. M. Mindlin 

State Technical University of Nizhny Novgorod, Russia 

E-mail address: ilia.mindlin@gmail.com
\bigskip

This paper addresses deep-water gravity waves of finite amplitude generated by an initial 
disturbance to the water. It is assumed that the horizontal 
dimensions of the initially disturbed body of the water are much larger 
than the magnitude of the free surface displacement in the origin of the waves. 

Initially the free surface has not yet been displaced from its equilibrium 
position, but the velocity field has already become different from zero. 
This means that the water at rest initially is set in motion suddenly 
by an impulse.

Duration of formation of the wave origin and the maximum water elevation 
in the origin are estimated using the arrival times of the waves and 
  the maximum wave-heights  at certain locations obtained from gauge records 
at the locations, and the distances between the centre of the origin and each 
of the locations. 

For points situated at a long distance from the wave origin, forecast is made 
for the travel time and wave height at the points. The forecast is based on 
the data recorded by the gauges at the locations.
\bigskip
 
{\bf 1. Problem outline. }
\bigskip                      

In the present study the deep-water waves are considered
which start to propagate away from an initially disturbed body of water. 
Then the water is acted on by no external force other than gravity. 
It is assumed that the free surface of the water is infinite in extent and 
the pressure along the surface is constant.
It is also assumed that the horizontal dimensions of the initially 
disturbed body of water are much larger than the maximum water elevation 
in the wave origin.

Below the case of plane waves is discussed.
 A sketch of the flow is shown in figure 1, where $(x,y)$-plane is the plane of flow, 
the $x$-axis is oriented upward and the $y$-axis in horizontal
direction. Let the curve $\,\,\Gamma\,\,$ be the trace of the free
surface $\,\,S\,\,$ in the $\,\,(x,y)\,\,$ plane,
$\,x=f<0,\,$ $y=0$ be the coordinates of the pole $\,O_1\,$ of the polar
coordinate system in the $\,(x,y)\,$ plane,
$\,\theta\,$ be the polar angle measured from the positive $x$-axis
in the counterclockwise direction, $\,t\,$ be the time. 
The external pressure, $\,P_*\,$, on the free surface is constant.
The liquid fills the space below the free surface.
The equilibrium position of the free surface is in horizontal plane $x=0$ 

Equations to the problem are presented in [1] and [2]. 

We begin by summarizing very briefly the pertinent theoretical results 
 that we need. The results for plane waves are proved, for instance, in [2].


\begin{figure}
	\resizebox{\textwidth}{!}
		{\includegraphics{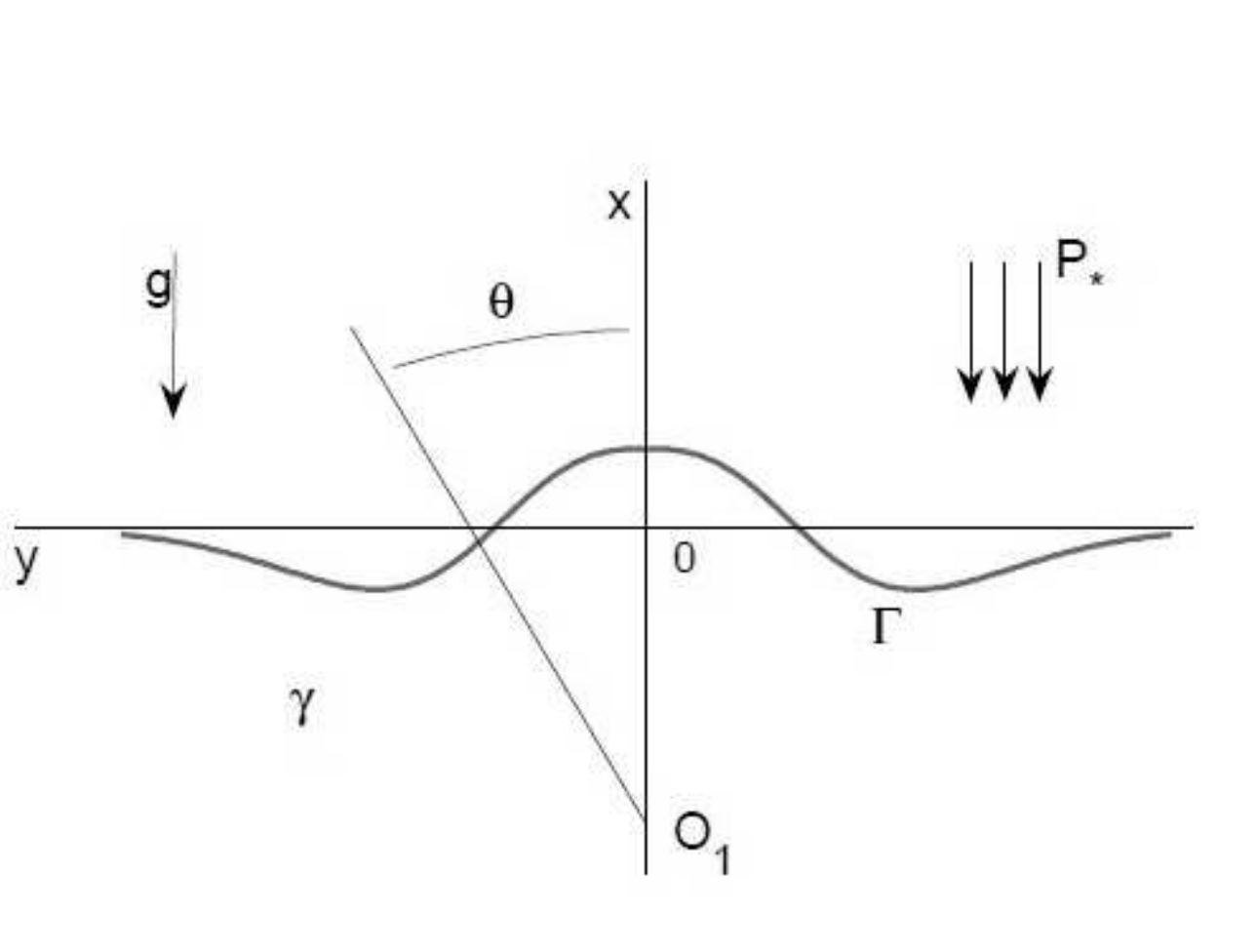}} 
	\caption{
	Coordinate systems and sketch of the free surface of a liquid. 
	}
	\label{qu-.1-.10}
\end{figure} 
\bigskip

{\bf 2. Preliminaries }
\bigskip 

{\bf 2.1. The equations of the free surface}
\bigskip  


The equations of the free surface are obtained in parametric form as follows
$$ 
x=cW_0(\theta,t),\\\ y=(x-f)\tan\theta, \\\
-\pi/2< \theta < \pi/2                                      \eqno(2.1.1)     
	$$         
$$
W_0(\theta,t)= \sum_{n=1}^{+\infty} (-1)^{n-1}\frac{1}{2n}
[a_nI_{2n+1}(\tau,\theta)+b_nJ_{2n+1}(\tau,\theta)]+                 
        $$
                        	$$          \eqno(2.1.2) $$   
$$
\frac{1}{\sqrt{2|f|}}\sum_{n=1}^{+\infty} (-1)^n
[\rho_nP_{2n}(\tau,\theta)+e_nQ_{2n}(\tau,\theta)],\,\,\,\,\,\
t=\tau\,{\sqrt{2|f|}}  
      $$
$$
I_{2n+1}(\tau,\theta)=\int\limits_0^{+\infty}
x^{2n+1}e^{-x^2/2}\cos(\frac{1}{2}x^2\tan\theta)\cos(\tau x)\,dx; \eqno(2.1.3) 
	$$
$$
J_{2n+1}(\tau,\theta)=\int\limits_0^{+\infty}
x^{2n+1}e^{-x^2/2}\sin(\frac{1}{2}x^2\tan\theta)\cos(\tau x)\,dx;  \eqno(2.1.4)
	$$
$$
P_{2n}(\tau,\theta)=\int\limits_0^{+\infty}
x^{2n}e^{-x^2/2}\cos(\frac{1}{2}x^2\tan\theta)\sin(\tau x)\,dx; \eqno(2.1.5)   
	$$
$$
Q_{2n}(\tau,\theta)=\int\limits_0^{+\infty}
x^{2n}e^{-x^2/2}\sin(\frac{1}{2}x^2\tan\theta)\sin(\tau x)\,dx. \eqno(2.1.6)   
	$$

The constants $a_n$, $b_n$ determine the initial displacement to the free surface, 
 the initial velocity field is determined by the constants $\rho_n$, $e_n$. 
The constants are independent of $f$.

Formally, equations (2.1.1) for each specified function
$\,\,W_0\,\,$ describe a family of curves depending on $\,f,\,$ $\,t\,$
 being considered as constant. The value  of $\,\,f\,\,$ 
determines the horizontal scale of the problem. 

Let $c$ be the maximum of the free surface displacement 
in the wave origin, so $\,\varepsilon=c/f$ is the ratio
of the maximum displacement to the characteristic horizontal dimension of the origin.
The problem was solved approximately up to small terms 
of order of $\,\varepsilon=c/f$.
\bigskip

{\bf 2.2.  Zeros of the wave group and terminology}
\bigskip     

Zeros of the wave group (2.1.1) are defined by the equation $W_0=0$ and 
(at fixed value of $\tau$) are situated in the numbered rays 
$\theta=\theta_k(\tau)$ ($k$ is the number of a zero). 
For brevity we will use the term 'zero $\theta(\tau)$' to denote a zero 
of a wave.
Zeros $\theta_k(\tau)$ of the wave group are independent of $f$. 

The term 'wave'\, means a
wave-group's section which is singled out by three zeros and consists 
of a crest and the trough following or preceding the crest.
The level difference between the crest and the trough  
is referred to as the wave height (or height of the wave), 
and the distances between two successive zeros is "half-wave-length". 

At any particular moment of time, each specific wave packet contains only one 
wave of maximum height (WMH) on semiaxis $y>0$  
(the situation with two waves of equal maximum height can be ignored), so
the zeros of WMH constitute a 'natural frame of reference' for other zeros:
let the zero $\theta_1$ be the front of the WMH, the zero $\theta_{-1}$ be 
the rear of the WMH, and the zero $\theta_0$ be between the  crest and the trough 
of the WMH; the distance between zeros $\theta_{-1}$ and $\theta_1$ is the length $l$ of the WMH: $l=|f\tan\theta_1-f\tan\theta_{-1}|$.
Calculations show that 
$$
|\tan\theta_1-\tan\theta_{-1}|\approx 5\,\,\,\,\,\,\hbox{for}\,\,\,\,\, 30<\tau<640.  \eqno(2.2.1)
	$$ 
 
All equations are written in non-dimensional variables.
Since the problem has no characteristic linear size, 
the dimensional unit of length, $\,\,L_*,\,\,$ is a free parameter.
But in the section 5, the value of $L_*$,   
as well as the value of $|f|L_*$, will be obtained  from instrumental data.
The dimensional unit of time, $\,\,T_*,\,\,$ is defined
by the relation $\,\,T_*^2g=L_*\,$, where $\,\,g\,\,$ is the acceleration
of free fall. The non-dimensional acceleration of free fall is equal to unity.
All parameters, variables and equations are made non-dimensional by 
the quantities $\,L_*,\,T_*,\,P_* $ and 
the density of water $\gamma=1000\,$ $\hbox{kg/m}^3$.
\bigskip

{\bf 2.3.  Specific wave packets}
\bigskip     

Spescific wave packets are defined by equations 
$$ 
x=I_{2n+1},\,\,\,\, y=(x-f)\tan\theta             \eqno(2.3.1)
	$$
$$
I_{2n+1}(\tau,\theta)=\int\limits_0^{+\infty}
x^{2n+1}e^{-x^2/2}\cos(\frac{1}{2}x^2\tan\theta)\cos(\tau x)\,dx;    
	$$
The subscript $2n+1$ is referred to as the packet number. 

In [2] the sequence of instantaneous forms of specific packets number 5 , 23, and the mixture of the two packets at $\tau = 25,\,50,\,100,\,200,\,$ and $300$ is shown.
We can see from the sequence of forms that the greater the packet number, the shorter the wavelength of its WMH, the slower the packet travels, the slower the packet disperses. 
It is clear from the sequence of instantaneous forms that the packet is growing in length. Three parts of the packet are observed on the surface, namely, the leading part of small amplitude which is followed by a middle part of larger amplitude; 
the "tail" of the packet is relatively short

Equations of other three sets of wave packets are obtained from 
(2.1.1), (2.1.2), (2.1.4) - (2.1.6). 

By equations (2.1.1) - (2.1.6), any wave group on the free 
surface is a mixture of finite or infinite (it depends on initial conditions) 
set of the specific wave packets of different numbers,  and evolution of each packet 
in the mixture is not influenced by evolution of the others. 
\bigskip

{\bf 3. Gauge records and data extracted from the records} 
\bigskip

In next sections a procedure is demonstrated for estimating some parameters 
of waves using arrival times of the waves of maximum height and their heights 
obtained from gauge records at some locations. To be assured 
that the horizontal dimensions of the wave origin are much larger than 
the magnitude of the free surface displacement in the origin, we use tsunami records at DART buoys deployed in the Pacific Ocean at a depth of 5000 metres. Records of the buoys and their locations can be found on the USA 
National Data Buoy Center public website (http://www.ndbc.noaa.gov/dart.shtml).

Figure 2 shows the Peruvian 08/2007 tsunami records de-tided 
using low-pass Buttherworth filter with 150 min cut-off; the residuals are 
non-tidal components, i.e., the mixture of seismic signal (of shorter period) 
and the tsunami waves.
\begin{figure}
	\resizebox{\textwidth}{!}                                              
		{\includegraphics{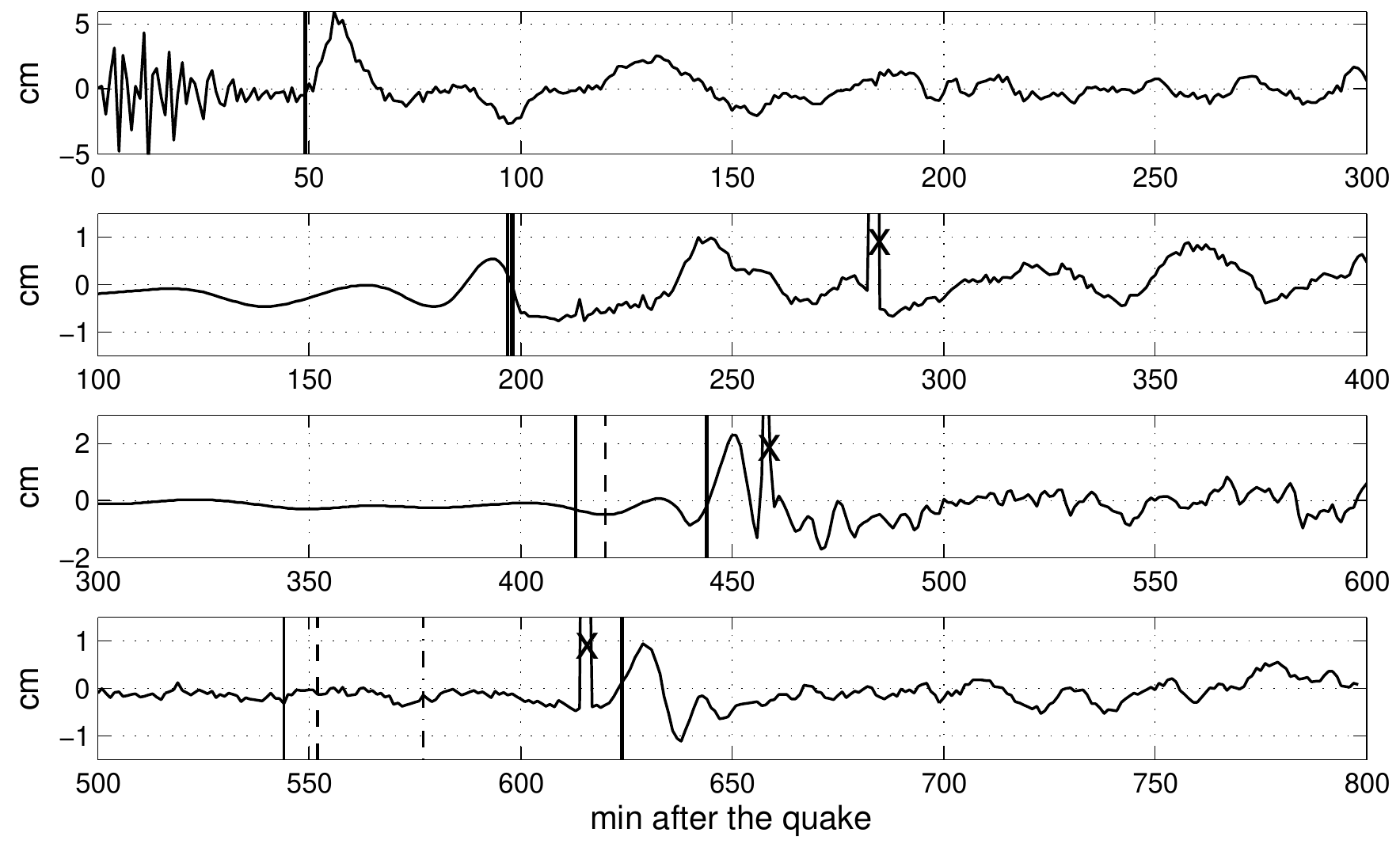}} 
	\caption{
De-tided records of the Peruvian 08/2007 tsunami at DART buoys (top to bottom) 	
32401, 32411, 51406, 46412. Vertical lines mark the arrival of the front of 
the WMH (thick solid line), its estimate with the first buoy record for the 
next three buoys (thin solid), its estimate with the first and second buoy 
records for the next two buoys (dashed), its estimate with the three buoys 
for the last one (dashdot). Crosses mark trigger pulses (signals send by an 
operator).
	}
\end{figure}  
\bigskip 

Data extracted from the records at the DARTS  
are summarized in Tables 1 - 3, where the travel time of the MHW, $t_*\,\hbox{min}$, 
the maximum wave-height, $H_*\,\hbox{cm}$, and
the distance between the gauge sensors and the centre (epicentre of the earthquake) of 
the wave origin, $y_*\,\hbox{km}$, are shown for each of the DARTs. 
\bigskip

\begin{center}
TABLE 1. The Peruvian tsunami: WMH as recorded at the DART buoys. 

Data 1.
\begin{footnotesize}
\begin{tabular}{|p{5mm} |p{15mm} |p{10mm} |p{10mm} |p{10mm} | p{15mm} | } 
\hline $\hbox{i}$ & $\hbox{Buoy}$ & $t_*\,\,\hbox{min.}$ & $H_*\,\,\hbox{cm} $ & 
  $y_*\,\, \hbox{km}$ & $y_*/t_*^2$ \\
\hline     1  & 32401    &     49     &    7.0    &    713   &  0.296960   \\
           2  & 32411    &     198    &    1.7    &    2561  &  0.065325   \\
           3  & 51406    &     444    &    3.9    &    5320  &  0.026986   \\
           4  & 46412    &     624    &    2.0    &    6921  &  0.017775   \\
\hline 
\end{tabular}\\
\end{footnotesize}
\end{center}
Figure 3 and Table 2 show the de-tided Dart records and data  for tsunami triggered by the earthquake near Kuril Islands in January 2007.

\begin{figure}
	\resizebox{\textwidth}{!}
		{\includegraphics{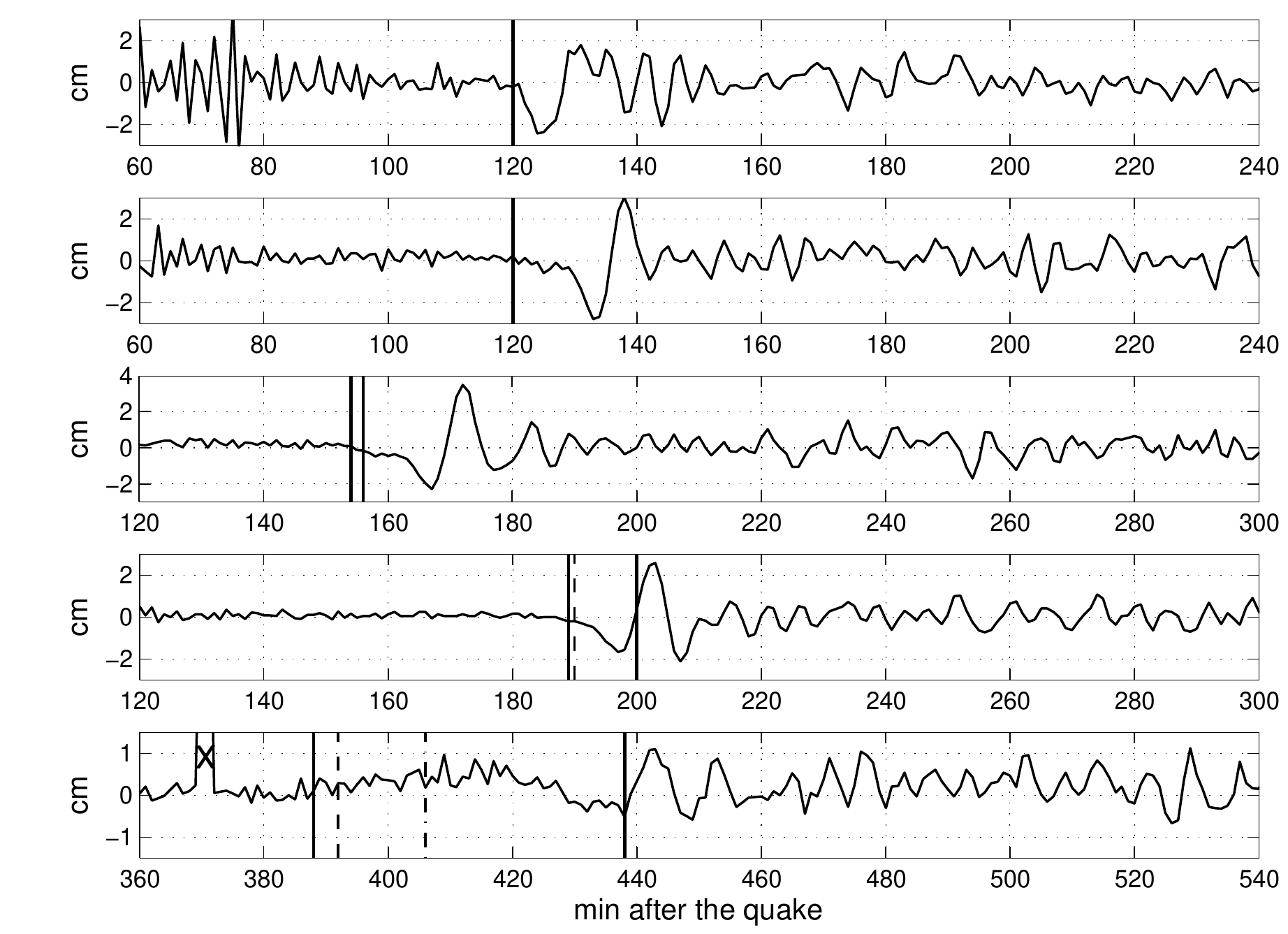} }  
	\caption{
De-tided records of the Kuril 01/2007 tsunami at DART buoys (top to bottom) 	
21413, 21414, 46413, 46408, 46419. Vertical lines mark the arrival of the front 
of the WMH (thick solid line), its estimate with the first and second buoy records 
for the next three buoys (thin solid), its estimate with the first three buoy records 
for the next two buoys (dashed), its estimate with the four buoys for the last one 
(dashdot). Cross marks a trigger pulse (signals send by an operator).
	}
\end{figure} 

\begin{center}
TABLE 2. The 2007 Kuril tsunami: WMH as recorded at the DART buoys.

Data 2.
 \begin{footnotesize}
\begin{tabular}{|p{15mm} |p{10mm} |p{10mm} |p{10mm} | p{15mm} | } 
\hline $\hbox{Buoy}$ & $t_*\,\,\hbox{min.}$ & $H_*\,\,\hbox{cm} $ & 
  $y_*\,\, \hbox{km}$ & $y_*/t_*^2$ \\
\hline     21413    &     120    &    4.0    &    1762  &  0.122361   \\
           21414    &     120    &    5.7    &    1804  &  0.125278   \\
           46413    &     154    &    5.7    &    2253  &  0.094999   \\
           46408    &     200    &    4.5    &    2660  &  0.066500   \\
           46419    &     438    &    1.6    &    5470  &  0.028513   \\
\hline 
\end{tabular}\\
\end{footnotesize}
\end{center}

 Tsunami was generated by the 
the earthquake occured near Kuril Islands in November 2006. 

The data obtained from the DART records of the tsunami are shown in Table 3.

\begin{center}
TABLE 3. The 2006 Kuril tsunami: WMH as recorded at the DART buoys. 

Data 3.
\begin{footnotesize}
\begin{tabular}{|p{15mm} |p{10mm} |p{10mm} |p{10mm} | p{15mm} | } 
\hline $\hbox{Buoy}$ & $t_*\,\,\hbox{min.}$ & $H_*\,\,\hbox{cm} $ & 
  $y_*\,\, \hbox{km}$ & $y_*/t_*^2$ \\
\hline     46413    &     155    &    8.5    &    2331  &  0.097024   \\
           46408    &     238    &    7.5    &    2735  &  0.048284   \\
           46402    &     270    &    10.0   &    3127  &  0.042894   \\
           46403    &     365    &    7.5    &    3581  &  0.026879   \\
\hline 
\end{tabular}\\
\end{footnotesize}
\end{center}
\bigskip

{\bf 4. Estimation of some parameters}
\bigskip 

{\bf 4.1. Model for the waves recorded}
\bigskip 

Assume the surface waves are described by equations 
$$ 
x=c\frac{1}{\sqrt{2|f|}}P_{2},\,\,\,\, y=(x-f)\tan\theta             \eqno(4.1.1)
	$$
$$
P_{2}(\tau,\theta)=\int\limits_0^{+\infty}
x^2e^{-x^2/2}\cos(\frac{1}{2}x^2\tan\theta)\sin(\tau x)\,dx.    
	$$
By the moment $t=0$ the free surface has not yet been displaced 
from its mean level (the horizontal plane $x=0$), but the velocity field has 
already become different from zero. 
This means that the motion of a body of water is triggerred by a sudden change in the velocity field. 
The wave packet (4.1.1) travels faster then any other specific packet.
\bigskip

{\bf 4.2. Properties of the specific wave packet (4.1.1)}
\bigskip

Horizontal coordinate of the zero $\theta_k(\tau)$ is given 
by $y_k=-f\cdot \tan\theta_k(\tau)$ and the function 
$P_2(\tau;\theta)$ is 
independent of $f$. This leads to the following 

{\it Assertion:} At any given value of $\tau$

i) for any wave of the packet the quantity $\Delta(\tau)=h(\tau)\sqrt{2|f|}$
($h(\tau)$ is the height of the wave) is independent of $f$; 

ii) the ratio of the distances $y_{k+1}(\tau)-y_k(\tau)$ and
$y_k(\tau)-y_{k-1}(\tau)$ between any successive zeros of the wave packet  
(and, consequently, the ratio of the lengths of two successeive waves) is
independent of $f$.

Let $L_*\,\,$ be the dimensional unit of length (in metres), then $T_*=\sqrt{L_*/g}\,\,$   is the dimensional unit of time (in seconds).

At an instant $t_*$ the dimensional coordinate of any zero 
$\theta=\theta(\tau)$ is 
$$
y_*(t_*)=|f|L_*\cdot \tan\theta(\tau)\cdot 10^{-3}\,\,\hbox{km},\,\,\,\,\,\,
t_*=\tau\sqrt{2|f|}\cdot T_*/60\,\,\hbox{min},
	$$
and, consequently, the ratio 
$$
\lambda(\tau)=\frac{y_*(t_*)}{t_*^2}\,\hbox{km/min$^2$}=
17.64\,\frac{\tan\theta(\tau)}{\tau^2}\,\hbox{km/min$^2$}.      \eqno(4.2.1)
	$$
depends on $\tau$ only. 

 Given a fixed value of $\tau$, $\tan\theta(\tau)$  can be calculated from equations 
(4.1.1) of the wave packet, and corresponding value of $\lambda(\tau)$ can be obtained 
from (4.2.1). 

Let $\theta_1(\tau)$ denote one of the three zeros $\theta(\tau)$ of the WMH,
which corresponds to maximum of the three $|\tan\theta(\tau)|$. 
By the zero $\theta_1(\tau)$ we define the front of WMH (see subsection 2.2). 
 
Table 4 shows computed characteristics of the WMH:
$\tan\theta_1,\,$ $\lambda(\tau),\,$ and 
$\Delta(\tau)=H(\tau)\sqrt{2|f|}\,\,$ ($\,\,cH(\tau)$ is the maximum wave height)
obtained from (4.1.1) and (4.2.1) (at $c=1$ $f=-10$).

\begin{center}
TABLE 4. Computed characteristics of the wave of maximum height.
\end{center}
\begin{footnotesize}
\begin{center}
\begin{tabular}{|p{10mm}  |p{10mm}  |p{10mm}  |p{10mm}  |p{10mm}  | p{10mm} | p{10mm}| } 
\hline $\tau$      & 25      & 30      & 35      & 40      & 45      & 50     \\ 
 $\tan\theta_1  $    & 19.032  & 19.818  & 23.687  & 27.587  & 31.508  & 32.531 \\                           
 $\lambda$         & 0.5371  & 0.3884  & 0.3411  & 0.3041  & 0.2744  & 0.2295  \\
 $\Delta(\tau) $         & 0.6326  & 0.5927  & 0.5653  & 0.5380  & 0.5126  & 0.4911  \\

$\lambda\cdot\tau$ & 13.429  & 11.653  & 11.938  & 12.166  & 12.351  & 11.477 \\

 $u $              &         & 0.1572  & 0.7738  & 0.7800  & 0.7842  & 0.2046  \\
\hline
\end{tabular}\\
\end{center}

\begin{center}
\begin{tabular}{|p{10mm}  |p{10mm}  |p{10mm}  |p{10mm}  |p{10mm}  | p{10mm} | p{10mm}| }
\hline $\tau$      & 55      & 60      & 65     & 70     & 75     & 80     \\ 
 $\tan\theta_1  $    & 36.396  & 40.268  & 44.159 & 48.057 & 49.114 & 52.974 \\                           
 $\lambda$         & 0.2122  & 0.1973  & 0.1844 & 0.1730 & 0.1540 & 0.1460 \\
 $\Delta(\tau) $         & 0.4738  & 0.4577  & 0.4420 & 0.4267 & 0.4155 & 0.4035 \\

$\lambda\cdot\tau$ & 11.673  & 11.839  & 11.984 & 12.110 & 11.552 & 11.681 \\

 $u $              & 0.7730  & 0.7744  & 0.7782 & 0.7796 & 0.2114 & 0.7720  \\
\hline
\end{tabular}\\
\end{center}

\begin{center}
\begin{tabular}{|p{10mm}  |p{10mm}  |p{10mm}  |p{10mm}  |p{10mm}  | p{10mm} | p{10mm}| }
\hline $\tau$      & 90     & 100    & 110    & 120    & 130    & 140    \\
 $\tan\theta_1  $    & 58.000 & 65.695 & 70.734 & 78.421 & 83.470 & 88.559 \\                           
 $\lambda$         & 0.1263 & 0.1159 & 0.1031 & 0.0960 & 0.0871 & 0.0797 \\
 $\Delta(\tau)  $        & 0.3840 & 0.3661 & 0.3507 & 0.3371 & 0.3255 & 0.3152 \\

$\lambda\cdot\tau$ & 11.368 & 11.589 & 11.343 & 11.528 & 11.326 & 11.158 \\

 $u $              & 0.5026 & 0.7695 & 0.5039 & 0.7687 & 0.5049 & 0.7606  \\
\hline
\end{tabular}\\
\end{center}

\begin{center}
\begin{tabular}{|p{10mm}  |p{10mm}  |p{10mm}  |p{10mm}  |p{10mm}  | p{10mm} | p{10mm}| }
\hline $\tau$      & 150    & 160     & 170     & 180     & 200     & 220      \\
 $\tan\theta_1  $    & 96.203 & 101.294 & 108.935 & 116.607 & 126.767 & 142.072  \\                           
 $\lambda$         & 0.0754 & 0.0698  & 0.0665  & 0.0635  & 0.0559  & 0.0518   \\
 $\Delta(\tau) $         & 0.3050 & 0.2960  & 0.2884  & 0.2804  & 0.2676  & 0.2554   \\

$\lambda\cdot\tau$ & 11.313 & 11.168  & 11.304  & 11.427  & 11.181  & 11.391   \\

 $u $              & 0.7644 & 0.5091  & 0.7641  & 0.7672  & 0.5081  & 0.7652  \\
\hline
\end{tabular}\\
\end{center}

\begin{center}
\begin{tabular}{|p{10mm}  |p{10mm}  |p{10mm}  |p{10mm}  |p{10mm}  | p{10mm} | p{10mm}| }
\hline $\tau$      & 240     & 260     & 280     & 300     & 310     & 320     \\
 $\tan\theta_1  $    & 154.802 & 167.532 & 180.265 & 192.996 & 198.071 & 205.727 \\ 
 $\lambda$         & 0.0474  & 0.0437  & 0.0406  & 0.0378  & 0.0364  & 0.0354  \\
 $\Delta(\tau) $         & 0.2446  & 0.2367  & 0.2281  & 0.2204  & 0.2175  & 0.2141  \\

$\lambda\cdot\tau$ & 11.378  & 11.357  & 11.357  & 11.348  & 11.271  & 11.341  \\

 $u $              & 0.6365  & 0.6366  & 0.6366  & 0.6365  & 0.5075  & 0.7656  \\
\hline
\end{tabular}\\
\end{center}

\begin{center}
\begin{tabular}{|p{10mm}  |p{10mm}  |p{10mm}  |p{10mm}  |p{10mm}  | p{10mm} | p{10mm}| }
\hline $\tau$      & 330     & 340     & 360     & 370     & 380     & 390    \\
 $\tan\theta_1  $    & 210.802 & 218.458 & 231.189 & 236.267 & 243.920 & 249.002 \\ 
 $\lambda$         & 0.0341  & 0.0333  & 0.0315  & 0.0304  & 0.0298  & 0.0289  \\
 $\Delta(\tau) $         & 0.2108  & 0.2077  & 0.2019  & 0.1991  & 0.1965  & 0.1950  \\

$\lambda\cdot\tau$ & 11.268  & 11.334  & 11.328  & 11.264  & 11.323  & 11.262  \\

 $u $              & 0.5075  & 0.7656  & 0.6365  & 0.5078  & 0.7653  & 0.5082  \\
\hline
\end{tabular}\\
\end{center}

\begin{center}
\begin{tabular}{|p{10mm}  |p{10mm}  |p{10mm}  |p{10mm}  |p{10mm}  | p{10mm} | p{10mm}| }
\hline $\tau$      & 400     & 410     & 420     & 440     & 460     & 470      \\
 $\tan\theta_1  $    & 254.094 & 261.734 & 269.387 & 279.559 & 292.291 & 299.931  \\                           
 $\lambda$         & 0.0280  & 0.0275  & 0.0269  & 0.0255  & 0.0244  & 0.0239   \\
 $\Delta(\tau) $         & 0.1926  & 0.1902  & 0.1879  & 0.1836  & 0.1796  & 0.1776   \\
                                                                               
$\lambda\cdot\tau$ & 11.205  & 11.261  & 11.314  & 11.208  & 11.209  & 11.257   \\

 $u $              & 0.5092  & 0.7640  & 0.7653  & 0.5086  & 0.6366  & 0.7640  \\
\hline
\end{tabular}\\
5\end{center}

\begin{center}
\begin{tabular}{|p{10mm}  |p{10mm}  |p{10mm}  |p{10mm}  |p{10mm}  | p{10mm} | p{10mm}| }
\hline $\tau$      & 480     & 490     & 500     & 520     & 540     & 550     \\
 $\tan\theta_1  $    & 305.023 & 312.663 & 317.762 & 330.494 & 343.227 & 350.867 \\                           
 $\lambda$         & 0.0233  & 0.0230  & 0.0224  & 0.0216  & 0.0208  & 0.0205  \\
 $\Delta(\tau) $         & 0.1758  & 0.1744  & 0.1729  & 0.1706  & 0.1674  & 0.1659  \\

$\lambda\cdot\tau$ & 11.209  & 11.256  & 11.211  & 11.211  & 11.212  & 11.253  \\

 $u $              & 0.5092  & 0.7640  & 0.5099  & 0.6366  & 0.6366  & 0.7640  \\
\hline
\end{tabular}\\
\end{center}

\begin{center}
\begin{tabular}{|p{10mm}  |p{10mm}  |p{10mm}  |p{10mm}  |p{10mm}  | p{10mm} | p{10mm}| }
\hline $\tau$      & 560     & 580     & 600     & 620     & 630     & 640     \\
 $\tan\theta_1  $    & 355.959 & 368.691 & 381.424 & 394.155 & 401.796 & 406.888 \\                           
 $\lambda$         & 0.0200  & 0.0193  & 0.0187  & 0.0181  & 0.0178  & 0.0175  \\
 $\Delta(\tau)$         & 0.1644  & 0.1616  & 0.1588  & 0.1563  & 0.1550  & 0.1538  \\

$\lambda\cdot\tau$ & 11.213  & 11.213  & 11.214  & 11.214  & 11.250  & 11.215  \\

 $u $              & 0.5092  & 0.6366  & 0.6366  & 0.6365  & 0.7641  & 0.5092  \\
\hline
\end{tabular}\\
\end{center}
\end{footnotesize}

In Table 4, values of the ratios $u=\Delta\tan\theta_1/\Delta\tau$ are given for each two neighbouring colomns (for instance, for the colomns $\tau=25$ and $\tau=30$ we find $u=(19.818-19.032)/(30-25)=0.1572$).

During a time interval $t_*$ min, the front of the WMH travels a distance 
$y_*(t_*)$ km at the average speed  
$$
V_*=\frac{y_*(t_*)}{t_*}=
\lambda(\tau)\cdot \tau \sqrt{2|f|}\cdot \frac{T_*}{60}\,\hbox{km/min}.   \eqno(4.2.2)  
	$$                                                            
The value of the front's average speed during time interval 
$\Delta \tau=\tau_{n+1}-\tau_n$ equals
$$
v_*=u\,\sqrt{|f|L_*g/2},\,\,\,\,\,\,              
u=[\tan\theta_1(\tau_{n+1})-\tan\theta_1(\tau_n)]/\Delta\tau,    \eqno(4.2.3)
$$
where $\tau_n$ and $u$ are given in Table 4. 

The values of $\lambda\tau$ and $u$ seems to suggest 
that the average speed of the front is nearly constant: 
in the interval $75\le\tau\le 640$, $\lambda\tau$ ranges from 11.158 to 11.680.  
\bigskip

{\bf 5. Estimation of parameters  using wave measurements} 
\bigskip

{\bf 5.1. Estimators for the wave model parameters}
\bigskip

For the front of the wave of maximum height (as for any zero) the following formulas hold 
$$
y_*(t_*)=-f\tan\theta_1(\tau)\cdot L_*>0,\,\,\,\,\,\,
t_*=\tau \sqrt{2|f|L_*/g}.
	$$

If at each locality $i$ data for the waves (4.1.1) were obtained from the records exactly,
the functions 
$$
S(a)=\sum_{i=1}^4\left(y_{*i}(t_{*i})-a\cdot \tan\theta_1(\tau_i)\right)^2\,\,\,\,\,\,
\hbox{and}\,\,\,\,\,\,F(a)= \sum_{i=1}^4 \left(60t_{*i}-\tau_i \sqrt{\frac{2a}{g}}\right)^2
	$$
would be equal to zero at $\,a=|f_*|$, where $|f_*|=|f|L_*$.  

But using equations (4.1.1) and values 
of $y_{*i},\,$ $t_{*i},\,$ $H_{*i}\,$ (obtained with errors) we estimate 
the value of $a$ by minimizing $S(a)$ or $F(a)$. 

Minimum value of $S(a)$ occurs at
$$
a=|f_{*1}|=\frac{ \sum_i {y_{*i}\tan\theta_{1i}}  }
{\sum_i {\tan^2\theta_{1i}}}\,\,\hbox{km},                    \eqno(5.1.1)      
	$$
while minimum value of $F(a)$ is reached at
$$
a=|f_{*2}|=\left(\frac{\sum_i {t_{*i}\tau_i}}{\sum_i {\tau^2_i}}\cdot 60\right)^2 4.9
\,\,\hbox{m}.                                                 \eqno(5.1.2)      
	$$
In the Table 4, the quantity $\Delta(\tau)=H(\tau)\sqrt{2|f|}$ is independent 
of $f$ ($cH(\tau)$ is the height of the WMH). 
If the data in Tables 1 - 3 were measured exactly at each locality $i$, the function 
$$
D(s)=\sum_i(c\Delta(\tau_i)\frac{1}{\sqrt{2|f|}}\,L_* -H_{*i})^2=
\sum_i\left(\Delta(\tau_i)cs-H_{*i}\right)^2,
\,\,\,\,\,\,cs=\frac{1}{\sqrt{2a}}\,L_*^{3/2}
	$$
would be equal to zero at true value of $s$.

At actual values of $H_{*i}$ the value of $cs$ is estimated by minimizing  
$D(s)$, which leads to the estimator 
$$
L_*^3=2a(c\eta)^2,\,\,\,\,\,\,c\eta=\frac{B}{A},\,\,\,\,\,\,  
A=\sum_i \Delta^2(\tau_i),\,\,\,\,\,\,
B=\sum_i \Delta_iH_{*i},                        \eqno(5.1.3)      
	$$
where $c\eta$ is independent of $c$ and $f$, $a$ is estimated by (5.1.1) or (5.1.2).
\bigskip

{\bf 5.2. Estimators for the wave parameters}
\bigskip

Calculations show that during some time interval, say $0\le\tau\le\tau_1$, 
a water hill in the form of a rounded solitary elevation symmetric with respect to
the vertical $x$-axis is appearing on the water surface. 
The height of the hill increases and reaches the maximum at $\tau=\tau_1=0.872$ 
On the interval $0\le\tau\le\tau_1$ there is only one zero 
$\theta(\tau)$ in the surface (at $y>0$), and the zero is almost immovable. Only at 
$\tau>\tau_1$ the heap of water begins to spread out and to turn into a wave group, 
which runs away from the vertical $x$-axis. The time interval $0\le\tau\le\tau_1$ is referred to as the interval of formation of the wave origin.

Calculations show that $\max \Delta(\tau)=\Delta(0.842)=1.036\,$ 
 (maximum is reached at $\theta=0$). This means that 
the maximum of the water elevation in the wave origin is given as
$$
h=c\Delta(0.842)\,L_*/\sqrt{2|f|}\,\,\,\,\hbox{or}\,\,\,h=1.036c\eta,                                   \eqno(5.2.1)    	
	$$
where $c\eta$ is given by (5.1.3).

Note that the right part of (5.2.1) is independent of $c$ and $f$.

Duration $t$ of the wave origin formation is estimated  
by ($g=9.8\,\,\hbox{m/s}^2$, $a$ is measured in metres)
$$
t=\tau_1 \sqrt{2|f|L_*/g}\,\,\,\,\hbox{or}\,\,\,\,t=0.842\sqrt{2a/g}\,\,\,\hbox{s}.                \eqno(5.2.2)  
	$$
By (2.2.1) length of the WMH is estimated as  
$$
l_*\approx 5a.                   \eqno(5.2.3)
	$$
Using formulas (4.2.2)   
 the average speed $V$ of the front of WMH during time interval $0<\tau<\tau_*$ may be estimated as 
$$
V_*=\lambda\cdot\tau_*\sqrt{\frac{2a}{g}}\cdot\frac{1}{60}\,\,\hbox{km/min}; \eqno(5.2.4)
	$$
formula (4.2.3) gives the estimate of the average speed during shorter time interval 
$$
v_*=u\sqrt{\frac{ag}{2}}.   \eqno(5.2.5)
	$$
\bigskip

{\bf 5.3. Theoretical characteristics of the WMH 
at locations of the DART buoys.  }
\bigskip

The Table 4 shows that $\tan\theta_1$ and $\lambda$ are monotone functions of $\tau$, so 
for given value of $\lambda$ the values of $\tau$ and $\tan\theta_1$ can be calculated 
using equations (4.1.1) or estimated using Table 4. 

For each DART location, setting $\lambda=y_*/t_*^2$ (see Tables 1 - 3) and using Table 4, 
we obtain the results, shown in Tables 5 - 7.

\begin{center}
TABLE 5. For Data 1: theoretical characteristics of the WMH  

at locations of the DART buoys. 

\begin{footnotesize}
\begin{tabular}{|p{15mm} |p{15mm} |p{15mm} |p{15mm} | p{15mm}| } 
\hline $\hbox{Dart}$ & $\lambda $ & $\tau $   & $\tan\theta_1(\tau)$ 
& $\Delta$ \\ 
\hline   32401     & 0.296960 & 41.206   &  28.533  & 0.531882  \\
         32411     & 0.065325 & 173.917  &  111.939 & 0.285267  \\
         51406     & 0.026986 & 418.567  &  268.290 & 0.188230  \\
         46412     & 0.017775 & 630.833  &  402.220 & 0.154900  \\
\hline
\end{tabular}\\
\end{footnotesize}
\end{center}
\bigskip

Values of $\tau,\,\,$ $\tan\theta_1(\tau),\,\,$ $\Delta(\tau)$ in the first line 
of Table 5
 are obtained with the use of $\lambda= 0.296960$ (in the first line of Table 1) and 
Table 4 as follows. 
We see from Table 4 that 
$0,2744<\lambda=0,296960<0,3041,\,\,$ $40<\tau<45,\,\,$ 
$27,587<\tan\theta_1(\tau)<31,508,\,\, $ $0,5126<\Delta(\tau)<0,5380$.

Method of linear interpolation gives $\tau=41,202,\,\,$ 
$\tan\theta_1(\tau)=28,530,\,\,$ $\Delta(\tau)=0,531894$. 
The rest lines in Table 5 are obtained in the same way. 

Tables 6  and 7 are similar to Table 5 and obtained in similar way.

\begin{center}
TABLE 6. 

For Data 2: theoretical characteristics of the WMH 

at locations of DART buoys.  

\begin{footnotesize}
\begin{tabular}{|p{15mm} |p{15mm} |p{15mm} |p{15mm} | p{15mm}| } 
\hline $\hbox{Dart}$ & $\lambda $ & $\tau $   & $\tan\theta_1(\tau)$ 
& $\Delta$ \\ 
\hline   21413     & 0.122361 & 93.787  &  60.914  & 0.374984 \\
         21414     & 0.125278 & 90.983  &  58.756  & 0.378995 \\
         46413     & 0.094999 & 121.125 &  78.987  & 0.335795 \\
         46408     & 0.066500 & 170.000 &  108.935 & 0.288400  \\
         46419     & 0.028513 & 394.300 &  251.192 & 0.193968 \\
\hline
\end{tabular}\\
\end{footnotesize}
\end{center}
\bigskip

\begin{center}
Table 7. 
For Data 3:  theoretical characteristics of the plane WMH 

at locations of the DART buoys. 

\begin{footnotesize} 
\begin{tabular}{|p{15mm} |p{15mm} |p{15mm} |p{15mm} | p{15mm}| } 
\hline $\hbox{Dart}$ & $\lambda $ & $\tau $   & $\tan\theta_1(\tau)$ 
& $\Delta$ \\ 
\hline   46413     & 0.097024 & 118.558  &  78.411  & 0.339047 \\
         46408     & 0.048284 & 235.888  &  154.791 & 0.246741 \\
         46402     & 0.042894 & 265.038  &  180.234 & 0.234534 \\
         46403     & 0.026879 & 420.300  &  279.545 & 0.187836 \\
\hline
\end{tabular}\\
\end{footnotesize}
\end{center}
\bigskip

{\bf 5.3. Applications of the estimators}
\bigskip

{\it Estimates based on Data 1.}

Table 1, Table 5, estimators (5.1.1) - (5.1.3) and (5.2.1) - (5.2.5) give 
$$
a_1=\frac{713\cdot28.533+2561\cdot111.939+5320\cdot268.290+6921\cdot402.220}{28.533^2+111.939^2+268.290^2+402.220^2}=18,284\,\,\,\hbox{km}
	$$
$$
a_2=\left(\frac{49\cdot41.206+198\cdot173.917+444\cdot418.567+624\cdot630.833}{41.206^2+173.917^2+418.567^2+630.833^2}\right)^2\cdot4.9=18278\,\,\,\hbox{m}
	$$
The maximum of the water elevation in the wave origin is 
$h=13.5\,\,\hbox{cm}.$ 

Duration $t$ of the wave origin formation is estimated  as 
$$
t_1=0.842\sqrt{2a_1/g}=51.43\,\,\,\hbox{s},\,\,\,\,\,             
t_2=0.842\sqrt{2a_2/g}=51.42\,\,\,\hbox{s}.
	$$
The WMH has the length 
$$
l_*\approx 91\,\,\,\,\hbox{km}.
	$$

It follows from the data of Table 1, that actual values of 
average speed $V_{*i}$ of the front of WMH (during time interval $0-t_*$) 
arrived at the buoys locations $i$ are respectively
$$
V_{*1}=\frac{713\cdot 60}{49}=873\,\hbox{km/h},\,\,\,
V_{*2}=776\,\hbox{km/h},\,\,\,V_{*3}=718\,\hbox{km/h},\,\,\,
V_{*4}=665\,\hbox{km/h}.
	$$
Using formula (5.2.4) and $a=|f_{*1}|=18284$ m, $g=9.8\,\hbox{m/s}^2$ we find 
$$
V_1=0.296960\cdot 41.206\cdot\sqrt{\frac{18284}{4.9}}\cdot\frac{1}{60}\,\hbox{km/min}=
12.46\,\hbox{km/min}=747\,\hbox{km/h},
	$$
$$
V_2=694\,\hbox{km/h},\,\,\,\,\,\,V_3=690\,\hbox{km/h},
\,\,\,\,\,\,V_4=685\,\hbox{km/h}.
	$$
From Tables 4 and 5, and formula (5.2.5) we find the estimate of the average speed during shorter time interval containing the point $t_i$. 
$$
v_i^*=u_i\sqrt{\frac{ag}{2}}
	$$
For buoy 32401 we have 
$40<\tau_1=41,168<45,\,$ $u=0,7842,\,$ $v_1^*=845\,\,\hbox{km/h}$. 
In the same way, for other three buoys we obtain  
$$
v_2^*=827\,\,\hbox{km/h}, \,\,\,\,v_3^*=826\,\,\hbox{km/h},\,\,\,\,
v_4^*=549\,\,\hbox{km/h}.
	$$
For the packet (4.1.1), the average speed of the wave of maximum height is 
estimated as 

$[5v_1^*+10(v_2^*+v_3^*+v_4^*)]/35=749\,\,\hbox{km/h}$.

The value of the parameter $\varepsilon =h/a$ is of order $10^{-5}$.
\bigskip

{\it Estimates based on Data 2.}

Using Table 2 and Table 6 we find
$$
a_1=23.256\,\,\,\hbox{km},\,\,\,\,\,\,a_2=23166\,\,\,\hbox{m}.
	$$
The maximum of the water elevation in the wave origin is 
$h=14.4\,\,\hbox{cm}$. 

Duration $t$ of the wave origin formation is estimated  as 
$$
t_1=58.00\,\,\,\hbox{s},\,\,\,\,\,             
t_2=57.89\,\,\,\hbox{s}.
	$$
The WMH has the length 
$$
l_*\approx 116\,\,\,\,\hbox{km}.
	$$

It follows from Table 2, that actual values of 
average speed $V_{*i}$ of the front of WMH (during time interval $0-t_*$) 
arrived at the buoys locations $i$ are respectively
$$
V_{*1}=881\,\hbox{km/h},\,\,\,
V_{*2}=902\,\hbox{km/h},\,\,\,V_{*3}=878\,\hbox{km/h},\,\,\,
V_{*4}=798\,\hbox{km/h},\,\,\,\,\,\,V_{*5}=749\,\hbox{km/h}.
	$$
Using formula (5.2.4) and $a=|f_{*1}|=23256$ m, $g=9.8\,\hbox{m/s}^2$ we find 
$$
V_1=790\,\hbox{km/h},\,\,\,\,\,\,	
V_2=785\,\hbox{km/h},\,\,\,\,\,\,V_3=793\,\hbox{km/h},
\,\,\,\,\,\,V_4=779\,\hbox{km/h},\,\,\,\,\,\,V_5=774\,\hbox{km/h}
	$$
\bigskip

{\it Estimates based on Data 3.}

Using Table 3 and Table 7 we find
$$
a_1=15.424\,\,\,\hbox{km},\,\,\,\,\,\,a_2=15.785\,\,\,\hbox{m}.
	$$
The maximum of the water elevation in the wave origin is 
$h=33.0\,\,\hbox{cm}$. 

Duration $t$ of the wave origin formation is estimated  as 
$$
t_1=47.24\,\,\,\hbox{s},\,\,\,\,\,             
t_2=47.79\,\,\,\hbox{s}.
	$$
The WMH has the length 
$$
l_*\approx 77\,\,\,\,\hbox{km}.
	$$

It follows from Table 3, that actual values of 
average speed $V_{*i}$ of the front of WMH (during time interval $0-t_*$) 
arrived at the buoys locations $i$ are respectively
$$
V_{*1}=902\,\hbox{km/h},\,\,\,
V_{*2}=689\,\hbox{km/h},\,\,\,V_{*3}=695\,\hbox{km/h},\,\,\,
V_{*4}=589\,\hbox{km/h}.
	$$
Using formula (5.2.4) and $a=|f_{*1}|=15424$ m, $g=9.8\,\hbox{m/s}^2$ we find 
$$
V_1=645\,\hbox{km/h},\,\,\,\,\,\,
V_2=639\,\hbox{km/h},\,\,\,\,\,\,V_3=638\,\hbox{km/h},
\,\,\,\,\,\,V_4=634\,\hbox{km/h}.
	$$
\bigskip

{\bf 6. Theoretical forecast for the waves recorded.}
\bigskip

For the waves (4.1.1), a line of forecasts of the WMH arrival time and amplitude at some buoys is produced corresponding to the timeline of the DART records.
\bigskip

{\bf 6.1. The forecast based on Data 1}

{\it The forecast based on one DART record.} 

Based on the measurements on DART buoy 32401, forecast of the travel time of 
the wave of maximum height that reaches each of the next three buoys and its height  
is prepared in the following way.

Using the first line of Table 1, the first line of Table 5, and estimators  (5.1.1) 
and (5.1.3) (it is supposed that the WMH has not reached the next three buoys yet) 
we obtain 
$$
a=|f_{*1}|=713/28.533=24.989 \,\,\,\hbox{km},\,\,\,\,\,\,c\eta=13.16\,\hbox{cm}.
	$$
For each of the next three buoys $\tan\theta_1=y_*/a$ is calculated, which is then used 
to  locate the values of $\tau$ and $\Delta$ between two appropriate consecutive values 
from Table 4. 

{\it Illustration.}  For the first line of Table 8  we find (for the buoy  32411) $\tan\theta_1=713/24.989=102.476.$ It follows from Table 4 that 
$101.294<102.476<108.935,\,\,\,$ $160<\tau<170,\,\,$ $0.2884<\Delta<0.2960$.
Method of linear interpolation gives $\tau=161.560,\,\,$  $\Delta(\tau)=0.294824$. 

Then, for each buoy, the travel time of the WMH and its height at the locations 
of the buoys are estimated by ($a$ is taken in metres) 
$$
t^*=\frac{\tau}{60}\cdot\sqrt{\frac{2a}{9.8}}\,\,\hbox{min},    \eqno(6.1.1)    
	$$
$$
H^*=|x_{max}-x_{min}|=c\frac{\Delta \cdot L_*}{\sqrt{2|f|}}=
c\Delta\sqrt{\frac{L_*^3}{2|f_*|}}=\Delta(\tau) \cdot c\eta.       \eqno(6.1.2)    
	$$
Note that the right hand side of (6.1.2) is independent of $c$ and $f$.

The forecast results are given in Table 8. 

\begin{center}
Table 8. For Data 1: the forecast based on one DART record.
\begin{footnotesize}
\begin{tabular}{|p{10mm} |p{10mm} |p{10mm} |p{10mm} | p{10mm} |p{10mm} |}  
\hline $\hbox{Buoy}$ & ${\tan}\theta_1 $ & $\tau $   & $\Delta$  
&$t^*\,\,\hbox{min}$ & $H^*\,\,\hbox{cm}$ \\ 
\hline   32411     & 102.486 & 161.560  &  0.294824 & 192 &  3.9  \\
         51406     & 212.896 & 332.734  &  0.209989 & 396 &  2.8   \\
         46412     & 277.364 & 435.685  &  0.184529 & 518 &  2.4  \\
\hline
\end{tabular}\\
\end{footnotesize}
\end{center}

It should be noted that formulas (5.1.1), (5.1.3), (5.2.2), and (5.2.1) are used 
when the forecast is based on arbitrary number of records.

{\it The forecast based on two DART records.} The following results are based on 
the measurements on DART buoys 32401 and 32411. The forecast is produced 
for the next two buoys.

The first two lines of Table 1,
the first two lines of Table 5, 
and formulas (5.1.1) and (5.1.3) give 
$$
a=|f_{*1}|=23.072 \,\,\,\hbox{km},\,\,\,\,\,\,\eta=9.17\,\hbox{cm}.
	$$
On the same lines as Table 8 we obtained Table 9 
(in figure 9, $t^*$ is marked by dashed line).

\begin{center}
Table 9. For Data 1: the forecast based on two DART records.
\begin{footnotesize}
\begin{tabular}{|p{10mm} |p{10mm} |p{10mm} |p{10mm} | p{10mm} |p{10mm} |}  
\hline $\hbox{Buoy}$ & ${\tan}\theta_1 $ & $\tau $   & $\Delta$  
&$t^*\,\,\hbox{min}$ & $H^*\,\,\hbox{cm}$ \\ 
\hline  
         51406     & 231.233 & 360.086  &  0.201838 & 411  & 2.3  \\           
         46412     & 301.255 & 472.600  &  0.177166 & 540  & 2.0  \\
\hline
\end{tabular}\\
\end{footnotesize}
\end{center}

{\it The forecast based on three DART records.} The following forecast is based on 
the measurements on DART buoys 32401, 32411, and 51406. 

The first three lines of Table 1, the first three lines of Table 5,  
and formulas (5.1.1) and (5.1.3) give 
$$
a=|f_{*1}|=20.339 \,\,\,\hbox{km},\,\,\,\,\,\,\eta=9.97\,\hbox{cm}.
	$$
\begin{center}
Table 10. For Data 1: the forecast based on three DART records.
\begin{footnotesize}
\begin{tabular}{|p{10mm} |p{10mm} |p{10mm} |p{10mm} | p{10mm} |p{10mm} |}  
\hline $\hbox{Buoy}$ & ${\tan}\Theta $ & $\tau $   & $\Delta$  
&$t^*\,\,\hbox{min}$ & $H^*\,\,\hbox{cm}$ \\ 
\hline  
          46412     & 340.986 & 536.480  &  0.168011 & 576 & 2.1  \\
\hline
\end{tabular}\\
\end{footnotesize}
\end{center}
\bigskip

For Data 1, the results are summarized in Tables 11  where 
for each of the mentioned DART buoy's locations the forecast of travel
time $t_k^*$ (in minutes) of the wave of maximum height and its height $H_k^*$ 
(in centimetres) are shown; the subscript $k$ shows that the forecast is based 
on measurements obtained from $k$ buoys. 

The actual travel time $t_*$ and height $H_*$ are repeated from Table 1.

\begin{center}
Table 11. For Data 1: the forecasts based on the model (4.1.1).
\begin{footnotesize}
\begin{tabular}{|p{10mm} |p{8mm} |p{8mm} |p{8mm} | p{8mm}| p{8mm} | p{8mm} | p{8mm} | p{8mm} | } 
\hline $\hbox{Buoy}$ & $t_1^* $ & $t_2^* $   & $t_3^*$ & $t_*$ &$H_1^*$ & $H_2^*$ & $H_3^*$ & $H_*$  \\
\hline 
         32411     & 192 & -   &  -   & 198 & 3.9 & -   & -   & 1.7 \\
         51406     & 396 & 411 &  -   & 444 & 2.8 & 2.3 & -   & 3.9 \\
         46412     & 518 & 540 &  576 & 624 & 2.4 & 2.0 & 2.1 & 2.0 \\
\hline
\end{tabular}\\
\end{footnotesize}
\end{center}

{\bf 6.2. The forecast based on Data 2}

The method of Subsection 6.1 is applied to the waves recorded at a number of DART buoys. 
The data obtained from the records are shown in Table 2. The results of forecasting are presented in Table 12 and figure 3.

\begin{center}
Table 12.  For Data 2: the forecasts based on the model (4.1.1).
\begin{footnotesize}
\begin{tabular}{|p{10mm} |p{8mm} |p{8mm} |p{8mm} | p{8mm}| p{8mm} | p{8mm} | p{8mm} | p{8mm} | } 
\hline $\hbox{Buoy}$ & $t_2^* $ & $t_3^* $   & $t_4^*$ & $t_*$ &$H_2^*$ & $H_3^*$ & $H_4^*$ & $H_*$  \\
\hline 
         46413     & 151 & -   &  -   & 154 & 4,4 & -   & -   & 5,7 \\
         46408     & 183 & 184 &  -   & 200 & 4,0 & 4,4 & -   & 4,5 \\      
         46419     & 366 & 374 &  391 & 438 & 2,9 & 3,1 & 3,1 & 1,6 \\
\hline
\end{tabular}\\
\end{footnotesize}
\end{center}

{\bf 6.3.  The forecast based on Data 3}

The Data 3 obtained from the records are shown in Table 3. The results of forecasting are presented in Table 13.

\begin{center}
Table 13. For Data 3: the forecasts based on the model (4.1.1).
\begin{footnotesize}
\begin{tabular}{|p{10mm} |p{8mm} |p{8mm} |p{8mm} | p{8mm}| p{8mm} | p{8mm} | p{8mm} |
p{8mm} | p{8mm} | p{8mm} | } 
\hline $\hbox{Dart}$ & $t_1^* $ & $t_2^* $   & $t_3^*$ & $t_*$ &$H_1^*$ & $H_2^*$ & $H_3^*$ & $H_*$  \\
\hline 
         46408     & 187 & -   &  -   & 238 & 7.8 & -   & -   & 7.5 \\ 
         46402     & 211 & 257 &  -   & 270 & 7.3 & 6.6 & -   & 10.0 \\
         46403     & 243 & 295 &  307 & 365 & 6.9 & 6.2 & 6.8 & 7.5 \\
\hline
\end{tabular}\\
\end{footnotesize}
\end{center}
\bigskip

The forecast results show earlier arrivals of the WMH ($t^*<t_*$). 
For Data 1 the percentage errors equal $(1-192/198)\cdot 100\%=3.0\%$ 
for buoy 32411, $(1-411/444)\cdot 100\%=7.4\%$ for buoy 51406, and 
$(1-576/624)\cdot 100\%=7.8\%$ for buoy 46412. 

The estimates $H_i^*$ of the maximum wave heights at the buoys exhibit no 
regularity: at the buoys, some of the estimates are close to the actual values 
$H_*$, and some 
of the estimates deviate from $H_*$. The range of the deviations $|H_i^*-H_*|$
is from $0$ to $H_*$.
Maximum water elevation in the wave origin is $h=0.13\,$m for Data 1, $h=0.14\,$m 
for Data 2, and $h=0.33\,$m for Data 3. 
The value of the ratio $h/|a|$ is of order $10^{-6} - 10^{-5}$.
We are not aware of instrumental measurements of water elevation in a wave 
origin.
 
Duration $t$ of the wave origin formation is estimated 
as $t=51\,\hbox{s}$ for Data 1, $t=58\,\hbox{s}$ for Data 2, and $t=47\,\hbox{s}$ for 
Data 3.

We are not aware of any theoretical estimates of the duration of wave origin formation, 
but in the available literature analogous estimates are given for 
the earthquakes:
typical earthquake with elastic energy release lasts 10 s (Watts 2001),  
"The duration of the rupture process is about 100 s, unusually long for 
its size" (Satake 1995)  
(i.e., the size of the 1992 Nicaragua earthquake source). 
Mean duration of an earthquake rupture is about 23 s 
(Burimskaya,  Levin \& Soloviev 1981).
\bigskip

{\it In passing.}
By the linear theory, speed $c$ of harmonic waves, their wavelength $l$, and water depth $d$ are related as follows
$$
c^2=\frac{gl}{2\pi}\tanh\frac{2\pi d}{l}.
	$$
For Data 1 the length of the wave of maximum height is 91 km, its speed 
$V\approx 750 \,\,\,\hbox {km/h}=208\,\,\,\hbox{m/s}$. 

At this values we get 
$$
\tanh\frac{2\pi d}{l}=0,305640,\,\,\,\,\,\,
\frac{2\pi d}{l}=0,32,\,\,\,\,\,\,d=4,8 \,\,\,\hbox{km}.
	$$
\bigskip

{\bf REFERENCES}
\bigskip

1. Mindlin, I.M.: Integrodifferential Equations in Dynamics of Heavy     
Layered Liquid,  Nauka$\ast$Fizmatlit, Moscow, 304 p, (1996) (Russian).

2. Mindlin, I.M.: Deep-water gravity waves: nonlinear theory of wave groups.
arXiv:1406.1681v1 [physics.ao-ph], 30 p, (6 Jun 2014)

3. Satake, K.  Linear and nonlinear computations of the 1992 Nicaragua 
earthquake tsunami. Pure and Appl. Geophys. vol. 144,  3/4, 455-470, (1995).	

4. Watts, Ph. Some opportunities of the landslide tsunami hypothesis. 
 Science of tsunami hazards, vol.19, 3, 126-149, (2001).

5. Burimskaya, R.N.,  Levin, B.V. \& Soloviev, S.L.  
A kinematic criterion for tsunamigenity of an underwater earthquake. 
 Dokl. Akad. Nauk. SSSR. {\bf 261}, 6, 1325-1330, (1981) (Russian).

\end{document}